 \documentclass[11pt]{article}
\input{amssym.def}
\input{amssym}
\newtheorem{theorem}{Theorem}
\newtheorem{proposition}[theorem]{Proposition}
\newtheorem{lemma}[theorem]{Lemma}
\newtheorem{corollary}[theorem]{Corollary}
\newtheorem{definition}[theorem]{Definition}

\def\H{{\mathcal H}}
\def\P{{\mathcal P}}
\def\PT{{\mathcal {PT}}}
\def\R{{\mathcal R}}

\def\R{\Bbb R} 
\def\Z{\Bbb Z} 
\def\N{\Bbb N}
\def\T{{\mathcal  T}}  
\def\C{\Bbb C}

\def\Sc{Schr\"o\-din\-ger}

\def\la{\langle}
\def\be{\begin{equation}}
\def\ee{\end{equation}}
\def\bea{\begin{eqnarray}}
\def\eea{\end{eqnarray}}
\def\ra{\rangle}
\def\ds{\displaystyle}

\def\om{\omega}

\def\l{\ell}

\begin{document}
\baselineskip=19pt
\begin{center}
{\large\bf\sc $\PT$ symmetric non-selfadjoint operators,  diagonalizable and non-diagonalizable, with real discrete spectrum}
\end{center}
\vskip 13pt
\begin{center}
 Emanuela Caliceti,  Sandro Graffi
 \\
 {\small Dipartimento di Matematica, 
Universit\`a di Bologna, 40127 Bologna, Italy 
\footnote{caliceti@dm.unibo.it, graffi@dm.unibo.it}}
\\
Johannes Sj\"ostrand
\\{\small 
Centre de Math\'ematiques, \'Ecole Polytechnique, 91125 Palaiseau, France 
\footnote{johannes@math.polytechnique.fr} } 
\end{center}
\begin{abstract}
\noindent
Consider in $L^2(\R^d)$, $d\geq 1$,  the  operator family
$H(g):=H_0+igW$.  
$\ds H_0=\,a^\ast_1a_1+\ldots +a^\ast_da_d+d/2$ is the quantum harmonic 
oscillator with rational  frequencies , $W$ a $\P$ symmetric bounded potential, and $g$ a real coupling 
constant.  
We show that if  $|g|<\rho$, $\rho$ being an explicitly determined constant, the spectrum of $H(g)$ is 
real and discrete.   Moreover we show that  the ope\-rator $\ds H(g)=a^\ast_1 a_1+a^\ast_2a_2+ig a^\ast_2a_1$ has real discrete spectrum but is not diagonalizable. 
\end{abstract}
\vskip 1cm   
%
 
%

\section{Introduction}
\setcounter{equation}{0}%
\setcounter{theorem}{0}%
A basic fact underlying $\PT$-symmetric quantum mechanics (see e.g. [1-10]; $\P$ is the 
parity operation, and $\T$ the complex conjugation) is 
the existence of non self-adjoint, and not even normal, but $\PT$-symmetric  \Sc\ operators (a particular case of complex symmetric operators, as remarked in \cite{PGP}) which have 
fully  real spectrum. 

Two natural  mathematical questions arising in this context  are (i) the determination of conditions under which $\PT$-symmetry actually yields real spectrum (for results in this direction see e.g. \cite{Shin},\cite{Tateo}, \cite{DD}, \cite{CGS}, \cite{CG}) and (ii) the examination of whether or not this phenomenon can still be understood  in terms of self-adjoint spectral  theory; for example, it has been remarked that if a $\PT$-symmetric \Sc\ operator with real spectrum is diagonalizable, then it  is conjugate to a self-adjoint operator through a similarity map (see e.g. \cite{KS}, \cite{Mo}, \cite{We}).    Hence the question arises whether $\PT$-symmetric \Sc-ªtype operators with real spectrum are always diagonalizable. 

In this paper a contribution is given to both questions. First, we solve in the negative the second one. Namely, we give a very simple, explicit example of a $\PT$ symmetric operator, with purely real and discrete spectrum, which cannot be diagonalized because of occurrence of Jordan blocks. The example is the following \Sc\ operator, acting in a domain $D(P(g))\subset L^2(\R^2)$ to be specified later:
\be
\label{E}
H(g):=a^\ast_1a_1+a^\ast_2a_2+ig a^\ast_2 a_1+1, \qquad g\in\R
\ee 
Here $a_i, a^\ast_i$, $i=1,2$ are the standard destruction and creation operators of  two independent harmonic oscillators:
\be
\label{CA}
a_i=\frac{1}{\sqrt 2}\left(x_i+\frac{d}{dx_i}\right), \quad a_i^\ast=\frac{1}{\sqrt 2}\left(x_i-\frac{d}{dx_i}\right),
\ee
so that (\ref{E}) can be rewritten under the form
\be
\label{E1}
H(g)=\frac12\left[-\frac{d^2}{dx_1^2}+x^2_1\right]+\frac12\left[-\frac{d^2}{dx_2^2}+x^2_2\right] +ig 
\frac12\left(x_2-\frac{d}{dx_2}\right)\left(x_1+\frac{d}{dx_1}\right)
\ee 
which is manifestly invariant under the $\P\T$-operation $x_2\to -x_2$, $ig \to -ig$. 

Second, we identify a new class of non self-adoint,  $\PT$-symmetric operators with purely real spectrum in $L^2(\R^d)$, $d>1$.  To our knowledge,  this is the first example of such operators in dimension higher than one (a preliminary version of this result, without proofs, already appeared in \cite{Ca}). An example of an operator belonging to this class is represented by a perturbation of the harmonic oscillators in dimension higher than one, namely by the following \Sc\ operators:
\be
\label{E2}
H(g)=\frac12\sum_{k=1}^d\left[-\frac{d^2}{dx_k^2}+\om_k^2x^2_k\right]+ig W(x_1,\ldots,x_d)
\ee
Here $W\in L^\infty(\R^d)$,  $W(-x_1,\ldots,-x_d) =-W(x_1,\ldots,x_d)$, $|g|<\rho$, where $\rho>0$ is an explicitly estimated positive constant, and the frequencies $\om_k>0$ are rational multiples of a fixed frequency $\om>0$: $\ds \om_k=\frac{p_k}{q_k}\om$.  Here   $p_k\in\N,q_k\in\N: k=1,\ldots,d$ is a pair of relatively prime numbers, with both $p_k$   and  $q_k$ odd, $k=1,\ldots,n$.  When $d=2$,  $\ds \frac{\om_1}{\om_2} =\frac{p}{q}$ this result can be strenghtened: if  $\om_1/\om_2 =p/q$, the spectrum is real if and only if  $p$ and $q$ are both odd. 

The paper is organized as follows: in the next section we work out the example (\ref{E}) making use of the Bargmann re\-presentation, in Section 3 we establish the class of $\P\T$-symmetric operators with real spectrum by  exploiting the real nature of  Rayleigh-\Sc\ perturbation theory (for related work on spectrum of $\PT$-symmetric operators through perturbation theory, see \cite{BD}, \cite{BMW}), and in Section 4 we work out the example represented by the perturbation of the resonant harmonic oscillators proving the above statements. 

\section{A  non diagonaliza\-ble $\P\T$ sym\-metric opera\-tor with real discrete spectrum}
\setcounter{equation}{0}%
\setcounter{theorem}{0}%
Consider the operator  $H(g)$ whose action on its domain is specified  by (\ref{E}) or, equivalently, (\ref{E1}).   Denote $H_0$ the operator corresponding to the two-dimensional harmonic oscillator, namely:
\be
\label{HO}
H_0:=\frac12\left[-\frac{d^2}{dx_1^2}+x^2_1\right]+\frac12\left[-\frac{d^2}{dx_2^2}+x^2_2\right] , \;D(H_0)=D(-\Delta)\cap D(x_1^2+x_2^2)
\ee
It is immediately verified that $\ds Vu:=a^\ast_2 a_1u\in L^2$ if  $u\in D(H_0)$.  Therefore we can give the following
\begin{definition}
The operator family  $H(g):g\in\R$ in $L^2(\R^2)$ is  the operator $H(g)$ whose action is $H_0+ig V$ on the domain $D(H_0)$. 
\end{definition}
Then we have:
\begin{theorem}
\label{t1}
Consider  the operator family $H(g)$ defined above. Then, $\forall \,g\in\R$, $\ds |g|<2$:
\begin{enumerate}
\item $H(g)$ has discrete spectrum. 
\item   All eigenvalues of $H(g)$  are  $\lambda_m=m+1, m=0,1,2,\ldots$.
Each eigenvalue $\lambda_m$ has geometric multiplicity $1$ but algebraic multiplicity $m+1$.
\end{enumerate}
More precisely: 
for each $m$ there is an $m$-dimensional subspace ${\cal H}_m$ invariant 
under $H(g)$ such that we have the orthogonal decomposition 
$\ds L^2=\bigoplus_{m=0}^\infty {\cal H}_m$; if we denote  
$\tilde{H}_m:= H|_{{\cal H}_m}$ the restriction of $H(g)$ to ${\cal H}_m$, 
then $\ds H(g)=\bigoplus_{m=0}^\infty \tilde{H}_m$ and $\tilde{H}_m$ 
is represented by the $(m+1)\times (m+1)$ matrix: 
\begin{eqnarray}
\label{J}
\tilde{H}_m =(m+1) I_{(m+1)\times (m+1) }+ig D_m
\end{eqnarray}
Here  $D_m$ is  a nilpotent of order $m+1$. Explicitly:
\be
\label{D}
 D_m:=\left(\begin{array}{ccccc} 0 & \sqrt{m} &\cdot & \cdot  & 0
\\
0 & 0 & \sqrt{2(m-1)} &\cdot &  0
\\
\cdot & \cdot & \cdot &  \sqrt{3(m-2)}&  \cdot
\\
\cdot & \cdot & \cdot &\cdot &  \sqrt{m}
\\
0 & 0 & \cdot & \cdot &  0
\end{array}\right)\Longrightarrow D_m^{m+1}=0
\ee
\end{theorem} 
\vfill\eject\noindent
{\bf Remarks}
\begin{enumerate}
\item ${\rm Spec}(H(g))$ is thus real and independent of $g$.
\item
  Formula (\ref{J}) is the Jordan canonical form of  $\tilde{H}_m$. The algebraic muliplicity is $m+1$.  
Since $D_m\neq 0$, $\tilde{H}_m$ is not diagonalizable by definition  and, a fortiori, neither is $H(g)$.
\end{enumerate}
{\bf Proof of Assertion 1} 
\newline
The classical Hamiltonians corresponding to the operators  $H_0$ and  $H(g)$ represent their symbols, denoted $\sigma_0(x,\xi)$ and $\sigma_g(x,\xi)$, respectively:
\begin{eqnarray}
\sigma_0(x,\xi)&=&\frac12(\xi_1^2+\xi_2^2+x_1^2+x_2^2), 
\\
\sigma_g(x,\xi)&=&\sigma_0(x,\xi)+ig \widetilde{\sigma}(x,\xi), \quad 
\widetilde{\sigma}(x,\xi):=\frac{1}{2}(x_2-i\xi_2)(x_1+i\xi_1)
\end{eqnarray}
We have indeed (formally) $\sigma_0(x,-i\nabla_x)=H_0$,  $\sigma_g(x,-i\nabla_x)=H(g)$. 
Since $\sigma_0\to +\infty$ as $|\xi|+|x|\to +\infty$, by well known results (see e.g.\cite{RS}, \S XIII.14) it is enough to 
prove that $\forall |g|<g^\ast=2$, and $\forall\,(x,\xi)$ outside some fixed ball 
centered in the origin of $\R^4$:
\be
\label{D1}
0<(1-\frac12|g|)\sigma_0(x,\xi)\leq |\sigma_g (x,\xi)|
\ee
To see this, we estimate:
$$
\vert \widetilde{\sigma } \vert\le {1\over 2}\vert x_2-i\xi _2 \vert
\vert x_1+i\xi _1 \vert\le {1\over 4}(\vert x_2-i\xi _2 \vert^2+
\vert x_1+i\xi _1 \vert^2)={1\over 2}\sigma _0,
$$
and hence
$$
\vert \sigma _g \vert\ge \vert \sigma _0 \vert-\vert g \vert \vert \widetilde{\sigma } \vert \ge (1-{\vert g \vert\over 2})\sigma _0.
$$
This proves the inequality and hence the assertion.
\vskip 3pt
To prove the remaining assertions of the theorem we  make use of  the Bargmann representation \cite{Ba}. To this end, recall the general definition of the Bargmann transform $U_B$ (even though we shall  need it  
only for $d=2$):
\be
\label{B}
(U_Bu)(z):=f(z)=\frac{1}{(2\pi)^d}\int_{\R^d}e^{-z^2+2\sqrt2\la z,q\ra -q^2 }u(q)\,dq, \quad z\in\C^d
\ee
Let us recall the relevant properties of the Bargmann transformation. 
\begin{enumerate}
\item $U_B$  is a unitary map  between $L^2(\R^d)$ and ${\cal F}={\cal F}_d$, the  space of all  entire holomorphic functions $f(z):\C^d\to \C$  such that (here $z=x+iy$):
\be
\label{B1}
\|f(z)\|^2_F:=\int_{\R^{2d}}|f(z)|^2e^{-|z|^2}\,dx\,dy=\la f,f\ra_F <+\infty
\ee
where the scalar product $\la f,g\ra_{\cal F}$ in ${\cal F}_d$ is defined by
\be
\label{B2}
\la f,g\ra_{\cal F}= \int_{\R^{2d}}f(z)\overline{g(z)}e^{-|z|^2}\,dx\,dy
\ee
Namely, with $f(z):=(U_B u)(z)$: $\|f(z)\|_{\cal F}=\|u(q)\|_{L^2(\R^d)}$. 
\item
Let  $a^\ast_i$, $a_i$ be the destruction and creation operators in the variable $x_i$  defined as in (\ref{CA}). Let $N_i:=a^\ast_ia_i$ be the corresponding number operator, $i=1,\ldots,d$.   Denote $\ds N^{(d)}:=\sum_{i=1}^dN_i$ the total number operator. Then we have:
\be
\label{B3}
U_Ba^\ast_iU_B^{-1}=z_i, \quad U_Ba_iU_B^{-1}=\frac{\partial }{\partial z_i}, \quad U_BN_dU_B^{-1}=\sum_{i=1}^d\,z_i \frac{\partial }{\partial z_i}
\ee
so that $H_0=N^{(2)}+1$.
The above operators are defined in their maximal domain in ${\cal F}_d$. 
Moreover:
\begin{eqnarray}
\label{B4}
Q(g)&:=&U_B(H(g)-1)U_B^{-1}=U_B(N^{(2)}+ig a^\ast_2 a_1)U_B^{-1}=
\\
\nonumber 
&=&z_1 \frac{\partial }{\partial z_1}+z_2\frac{\partial }{\partial z_2}+ig z_2\frac{\partial}{\partial z_1}:=Q_0+ig W
\end{eqnarray}
defined on the maximal domain. Remark that ${\rm Spec}\,(Q_0)=\{0,1,\ldots,m,\ldots\}$. The eigenvalue $\lambda_m=m$ has multiplicity $m+1$.
\item Let $\psi_k(x)$ be the normalized eigenvectors of the one-dimensional harmonic oscillator in $L^2(\R)$. Then:
\be
\label{B5}
(U_B\psi_k)(z):=e_k(z)=\frac{1}{\sqrt {\pi^{1/2} k!}}z^k, \quad k=0,1,\ldots
\ee
 \end{enumerate}
 Let now $m=0,1,2,\ldots$. Define:
\begin{eqnarray*}
f_{m,h}(z_1,z_2)&:=&e_{m-h}(z_2)e_h(z_1),\;h=0,\ldots,m;
\\
 {\cal K}_m&:=&{\rm Span}\{f_{m,h}: h=0,\ldots,m\}=
 \\
 &=& {\rm Span}\{e_{l_1}(z_2)e_{l_2}(z_1):l_1+l_2=m\}
\end{eqnarray*}
Hence the following properties are immediately checked:
\be
\label{B61}
{\rm dim}\, {\cal K}_m=m+1; \quad  {\cal K}_m\perp {\cal K}_l, \;m\neq l;\quad \bigoplus_{m=0}^\infty {\cal K}_m={\cal F}_2
\ee
We then have
\begin{lemma}
\label{l1} 
\vskip 1pt\noindent
\begin{enumerate}
\item 
For any $m=0, 1,\ldots$:
\be
\label{B7}
Q(g)f_{m,h}=mf_{m,h}+ig h f_{m,h-1},\quad h=0,\ldots,m .
\ee
\item Let  $\Pi_m$ be the orthogonal projection  from ${\cal F}_2$ onto ${\cal K}_m$.
Then: 
\newline
$[\Pi_m,Q(g)]=0$; equivalently,  ${\cal K}_m$ reduces $Q(g)$: $Q(g){\cal K}_m\subset {\cal K}_m$; 
\item
Let $Q(g)_m:=Q(g)|_{{\cal K}_m}=\Pi_mQ(g)\Pi_m=\Pi_mQ(g)=Q(g)\Pi_m$ be ${\cal K}_m$-component  of $Q(g)$. 
Then $\ds Q(g)=\bigoplus_{m=0}^\infty Q(g)_m$;
\end{enumerate}
\end{lemma}
{\bf Proof}  
\newline
1. Just compute the action of $Q(g)$ on $f_{m,h}$:
\begin{eqnarray}
\nonumber
Q(g)f_{m,h}&=&(z_1 \frac{\partial }{\partial z_1}+z_2\frac{\partial }{\partial z_2}+ig z_2\frac{\partial}{\partial z_1})e_{m-h}(z_2)e_h(z_1)
\\
&=&
\label{B8}
(m-h)e_{m-h}(z_2)e_h(z_1)+he_{m-h}(z_2)e_h(z_1)
\\
\nonumber
&+&ig \sqrt{h(m-h+1)}e_{m-(h-1)}(z_2)e_{h-1}(z_1)
\\
\nonumber
&=&
mf_{m,h}+ig\sqrt{h(m-h+1)} f_{m,h-1}
\end{eqnarray}
2. Since the vectors $f_{m,h}:h=0,\ldots,m$ span  ${\cal K}_m$, by linearity the above formula  entails \linebreak $Q(g){\cal K}_m\subset {\cal K}_m$. 
\newline
3.  The assertion follows from 2. above and the completeness relation (\ref{B61}). 
\vskip 3pt\noindent
{\bf Proof of Theorem \ref{t1}}   
\newline
We have to prove Assertion 2.
\newline
2.  Making $h=0$ in (\ref{B8}) we get:
$$
Q(g)f_{m,0}=mf_{m,0}, \quad m=0,1,\ldots
$$
Hence $\lambda^\prime_m=m$ is an eigenvalue of $Q(g)$ with eigenvector $f_{m,0}$, i.e. with geometric multiplicity one. By  the unitary equivalence  $H(g)=U_B^{-1}(Q(g)+1)U_B$ we conclude that $\lambda_m=m+1, m=0,\ldots,$ is an
eigenvalue of $H(g)$ of geometric multiplicity one, with eigenvector 
$U_B^{-1}f_{m,0}=\psi_m(x_1)\psi_0(x_2)$.
From (\ref{B7}) we read off the matrix representation (\ref{J}), 
(\ref{D})
and we get the statement about the algebraic multiplicity. On account of the unitary equivalence $ {\cal K}_m=U_B^{-1} {\cal H}_m$ this concludes the proof of the theorem.
\vskip 0.5cm\noindent
\section{A  class of non self-adjoint $\P\T$ symmetric ope\-rators with real  discrete spectrum}
\setcounter{equation}{0}%
\setcounter{theorem}{0}%
Let $H_0$ be a selfadjoint operator in $L^2(\R^d), d\geq1$, bounded below 
(without loss of generality, positive) with compact resolvent, and let 
$ D(H_0)$ denote its domain. Let $\P$ be the parity operator in $L^2(\R^d)$ defined by 
\be
\label{P}
(\P\psi)(x) = \psi(-x)\,,\quad \forall\psi\in L^2(\R^d)\,,\;\forall x\in\R^d\,.
\ee
Let us assume that $H_0$ is $\P$-symmetric, i.e. 
\be
\label{Psymm}
\P H\psi=H\P\psi\,, \quad \forall\psi\in D(H_0)
\ee
and also $\T$-symmetric, i.e.
\be
\label{Tsymm}
(\overline {H_0\psi})(x) = (H_0\overline{\psi})(x)\,,\quad\forall\psi\in D(H_0)\,,\;\forall x\in\R^d\,.
\ee
Let $0<\ell_1<\ell_2 <\ldots $ be the increasing sequence of the  eigenvalues of $\H_0$. Let $m_r$ denote the multiplicity of $\ell_r$ and $\psi_{r,s}, \,s=1,...,m_r$, denote $m_r$ linearly independent eigenfunctions corresponding to $\l_i$, which form a basis of the eigenspace 
\be
\label{Span}
{\cal M}_r:= {\rm Span}\{\psi_{r,s}: s=1,...,m_r\}
\ee
corresponding to $\ell_r$. 
\begin{definition}
\vskip 1pt\noindent
\begin{enumerate}
\item An eigenspace  ${\cal M}_r$ is even (odd) if all basis vectors $\{\psi_{r,s}:s=1,...,m_r\}$ are even (odd); i.e., if either $\P\psi_{r,s}= \psi_{r,s},  \forall s=1,\ldots,m_r$, or 
 $\P\psi_{r,s}= -\psi_{r,s},  \forall s=1,\ldots,m_r$.
\item An eigenvalue $\ell_r$ is even (odd) if the corresponding eigenspace   ${\cal M}_r$ is even (odd). 
\end{enumerate}
\end{definition}
\par\noindent
 Now, let $W \in L^{\infty}(\R^d)$  be an odd real function, i.e. $W(x) = -W(-x),\, \forall x\in\R^d$.  Let $V:= iW$; clearly $V$ is $\P\T$- even, i.e.
\be
\label{V}
\overline{V(-x)} = V(x)\,,\quad \forall x\in \R^d\,.
\ee
Then, $\forall g\in\C$, the operator $ H(g):= H_0 + gV$ defined on  $D(H(g)) = D(H_0)$ by
\be
\label{H(g)}
H(g)\psi = H_0\psi + gV\psi\,, \quad \forall \psi\in D(H_0)
\ee
is closed. More precisely $H(g)$ represents an analytic family of type A  of closed operators in the sense of Kato (\cite{Ka}, Ch. VII.2) for $g\in\C$,  
with compact resolvents.  Thus ${\rm Spec}(H(g))$ is discrete for all $g$. For $g\in\R$ the operator $H(g)$ is $\P\T$-symmetric, i.e.
\be
\label{PT}
\overline{\P H(g)\psi}(x) = H(g)\overline{\psi}(-x)\,, \quad \forall\psi\in D(H_0)\,.
\ee
Moreover:
\be
\label{agg}
H(g)^\ast=H(-g)
\ee
We want to prove the following result.
\begin{theorem} 
\label{thm2} Let $H_0$ and $W$ enjoy the above listed properties. 
Assume furthermore:
\begin{itemize} 
\item [(1)]  $\ds \delta:=\frac12\inf_{r}(\ell_{r+1}-\ell_r)> 0$;
\item[(2)] Each eigenvalue $\ell_r: r=1,\ldots$ is either even or odd.
\end{itemize}
Then if $\ds |g|<\frac{\delta}{\|W\|_\infty}$ each eigenvalue $\lambda(g)$ of $H(g)$ is real, and thus the spectrum of $H(g)$ is purely real. 
\end{theorem}
{\bf Example}
\newline
The $d$-dimensional harmonic oscillator with equal frequencies 
\be
\label{H_0}
H_0=\frac12\sum_{k=1}^d\left[-\frac{d^2}{dx_k^2}+\om^2x^2_k\right]
\ee
has the properties  required by $H_0$.  In this case indeed:
$$
\ell_r=\om(r_1+\ldots+r_d+d/2):=\om(r+d/2),\quad r_k=0,1,2,\ldots; \;k=1,\ldots,d
$$
with multiplicity $m_r=(r+1)^d$.  Here the corresponding eigenspace is: 
$$
{\cal M}_r:= {\rm Span}\{\psi_{r,s}:s=1,...,m_r\}= {\rm Span}\{\psi_{r_1}(x_1)\cdots \psi_{r_d}(x_d): r_1+\ldots+r_d=r\}
$$
where, as above, $\psi_r(x)$ is an Hermite function.   Now if $r$ is odd the sum  $r=r_1+\ldots+r_d$ contains an odd number of odd terms;  since $\psi_s(x)$ is an odd function when $s$ is odd,  the product $\psi_{r_1}(x_1)\cdots \psi_{r_d}(x_d)$ contains  an odd number of odd factors and  is therefore odd.  $\ell_r$ is therefore an odd eigenvalue. An analogous argument  shows that $\ell_r$ is an even eigenvalue when $r$ is even.  Moreover, $\ell_{r+1}-\ell_r=\omega$ and thus condition (1) above is fulfilled.
\vskip 3pt
Actually, the above example is a particular case of a more  general statement, while for  $d=2$ the above 
application to the perturbation of harmonic oscillators can be considerably strenghtened.
\begin{theorem}
\label{thm3}
Let 
\be
\label{2H_0}
H_0=\frac12\sum_{k=1}^d\left[-\frac{d^2}{dx_k^2}+\om_k^2x^2_k\right]
\ee
Assume the  frequencies to be  rational multiples of a fixed frequency $\om>0$, namely:
\be
\label{mul}
\om_k=\frac{p_k}{q_k}\om , \quad k=1,\ldots,d
\ee
where $(p_k,q_k)$ are relatively prime  natural numbers. Then:
\begin{itemize}
\item[(i)] If  $p_k$ and $q_k$ are both odd, $k=1,\ldots,d$, the assumptions of Theorem  \ref{thm2} are fulfilled; 
\item[(ii)] If $d=2$,  the condition (\ref{mul}) $p_k$ and $q_k$ both odd is also necessary for the validity of assumption (2) of   Theoren \ref{thm2}, while  assumption (1) holds independently of the parity of $p_k$, $q_k$.
\end{itemize}
\end{theorem}
We will now prove Theorem \ref{thm2} in two steps (Propositions \ref{propo1} and and \ref{anglobale}), while the proof of Theorem \ref{thm3} is postponed 
to the next Section.  In the first step we show that the degenerate Rayleigh-\Sc\ perturbation 
theory near each eigenvalue $\ell_r$ is real and convergent, 
with a convergence radius independent of $r$. Thus there exists $\rho>0$ such 
that all the $m_r$ eigenvalues  near $\ell_r$ (counted according to their multiplicity) 
existing for $|g|<\rho$ are real for all $r$. The second step is the proof that  
$H(g)$ admits no other eigenvalue for $|g|<\rho$. To formulate the first  step, we recall some relevant  notions and results of perturbation theory. 
\par
Let $g_0\in \C$ be fixed and let $\mu$ be an eigenvalue of $H(g_0)$. Let $c>0$ be sufficiently small so that 
$$
\Gamma _{c} = \{z:\mid z-\mu\mid = c\}
$$
encloses no other eigenvalue of $H(g_0)$. Then for $|g-g_0|$ small 
$\Gamma _{c}$ is contained in the resolvent set of $H(g)$, $\rho(H(g)):=
\C\setminus{\rm Spec}(H(g))$. Moreover $\Gamma _c\subset{\cal D}$, where 
\begin{eqnarray*}
&& {\cal D}:= \{z\in \C: \exists b(z)>0\; s.t. \;(z-H(g))^{-1}:=R_g(z)
\\
&& {\rm exists\, and \,is \,uniformly\, bounded\, for}\, |g-g_0|<
b(z)\}.
\end{eqnarray*}
Then for $|g-g_0|$ sufficiently small
\be
\label{Projection}
P(g) = (2\pi i)^{-1} \oint _{\Gamma _{c}} R_g(z)\,dz
\ee
is the projection corresponding to the part of the spectrum of $H(g)$ 
enclosed in $\Gamma _{c}$ and $\forall z\in{\cal D}$
\be
\label{NRC}
\|R_g(z) - R_{g_0}(z)\| \to 0\,,\quad {\rm as}\; g\to g_0
\ee
whence
\be
\label{NPC}
\|P(g) - P(g_0)\| \to 0\,,\quad {\rm as}\; g\to g_0
\ee
(see e.g.\cite{Ka}, \S VII.1).  In particular, if $m$ denotes the multiplicity 
of $\mu$, for $g$ close to $g_0$, $H(g)$ has exactly $m$ eigenvalues (counting multiplicity) 
inside $\Gamma _{c}$, denoted $\mu_{s}(g), s=1,...,m$, which converge to $\mu$ as $g\to g_0$. If we denote by 
${\cal 
M}(g)$ the range of the projection operator $P(g)$, then ${\rm dim}\,{\cal M}(g)=m$ as $g\to g_0$, and 
$H(g){\cal M}(g)\subset {\cal M}(g)$. Hence the component 
$P(g)H(g)P(g)=P(g)H(g)=H(g)P(g)$ of $H(g)$ in ${\cal M}(g)$ has rank 
$m$ and its  eigenvalues  are precisely $\mu_{s}(g), s=1,...,m$.   
\vskip 0.3cm\noindent
Assume from now on  $g_0=0$ so that the unperturbed operator is the  self-adjoint  operator $H_0:=H(0)$. Let $\ell= \ell_r, r=1,2,\dots$, be a fixed eigenvalue of $H_0$, $m=m_r$ its multiplicity and $\psi_s:= \psi_{r,s}: s=1,\ldots,m$ be an orthonormal basis in ${\cal M}_r:={\cal M}_r(0)$.  Then there is $\bar g(r)>0$ such that  the vectors $P_r(g)\psi_{r,s}: s=1,\ldots,m$ are a basis in the invariant subspace ${\cal M}_r(g)$ for $|g|<\bar g(r)$.  We denote $\phi_{r,s}(g): s=1,\ldots,m$ the orthonormal basis in ${\cal M}_r(g)$ obtained from  $P_r(g)\psi_{r,s}: s=1,\ldots,m_r$ through the Gram-Schmidt orthogonalization procedure. Then the eigenvalues  $\mu_s(g)=\l_{r,s}(g), s=1,...,m_r$, are the eigenvalues of the $m_r\times m_r$ matrix $T_{r}(g)$ given by:
\begin{eqnarray*}
(T_{r}(g))_{hk}:=\la\phi_{r,h}(g), H(g)P(g)\phi_{r,k}(g)\ra=\quad 
\\
\la\phi_{r,h}(g), P_r(g)H(g)P_r(g)\phi_{r,k}(g)\ra,\quad h,k=1,\ldots,m_r\,.
\end{eqnarray*}
Let $\ds \phi_{r,s}(g) = \sum_{j=1}^m \alpha_{sj}^r(g)P_r(g)\psi_{r,j}\,,\; \alpha_{sj}^r(g)\in\C\,, s,j=1,\dots,m_r$. Then
\be\label{matrixT}
(T_{r}(g))_{hk} = \sum_{j,l=1}^m
\alpha_{hj}^r(g)\overline{\alpha_{kl}^r(g)}
\langle\psi_{r,j}, P_r(-g)H(g)P_r(g)\psi_{r,l}\rangle \,, \quad h,k=1,\ldots,m\,.
\ee
Consider now the $m_r\times m_r$ matrix $B_r(g) = (B_{jl}^r(g))_{j,l=1,\dots,m}$, where 
\be\label{matrixB}
B_{jl}^r(g) = \langle \psi_{r,j}, P_r(-g)H(g)P_r(g)\psi_{r,l}\rangle \,, \quad j,l=1,\ldots,m_r\,.
\ee
Its self-adjointness entails the self-adjointness of $T_{r}(g)$. We have indeed:
\begin{lemma}
Let $B_{jl}^r(g) = \overline{B_{lj}^r(g)}, \forall j,l=1,\dots,m_r$. Then:
 $$
 (T_{r}(g))_{hk} = \overline{(T_{r}(g))_{kh}},\, h,k=1,\ldots,m_r.
 $$
\end{lemma}
{\bf Proof}
\newline
Since $B_{jl}^r(g) = \overline{B_{lj}^r(g)}, \forall j,l$ we can write:
\begin{eqnarray}
\label{Tselfadjoint}
\overline{(T_{r,}(g))_{kh}} &=& \overline{\sum_{p,s=1}^{m_r} \alpha_{kp}^r(g)\overline{ \alpha_{hs}^r(g)}B_{ps}^r(g)} = \sum_{p,s=1}^{m_r} \overline{\alpha_{kp}^r(g)} \alpha_{hs}^r(g)B_{sp}^r(g) 
\nonumber\\
&= &\sum_{j,l=1}^{m_r}\alpha_{hj}^r(g) \overline{\alpha_{kl}^r(g)}B_{jl}^r(g) = (T_{r}(g))_{hk}\,.
\end{eqnarray}
and this proves the assertion.

In other words the selfadjointness of $T_{r}(g))$, and thus the reality of the eigenvalues  $\ell_{r,s}(g)$ for $|g|<\bar g(r)$, follows from the selfadjointness of  $B_r(g)$ which will be proved by the construction of the Ray\-leigh-Sch\"o\-dinger perturbation expansion (RSPE) for the operator $P_r(-g)H(g)P_r(g)$, which we now briefly recall, following  (\cite{Ka}, \S II.2.7;  here $T^{(1)}=V=iW$, 
$T^{(\nu)}=0,$, $\nu\geq 2$, $D=0$).
\par\noindent
\begin{itemize}
\item[(1)] The geometric expansion in powers of $g$ of the resolvent  
$$
R_g(z)= (z-H(g))^{-1}=(z-H_0-gV)^{-1}=R_0(z)\sum_{n=0}^\infty(-g)^n[VR_0(z)]^n
$$ 
is norm convergent for $|g|$ suitably  small. Insertion in (\ref{Projection}) yields the expansion for $P(g)$:
\begin{eqnarray}
\label{exp0}
P_r(g)&=&\sum_{n=0}^\infty g^n P^{(n)}_r, \quad P^{(0)}_r=P_r(0):=P_r
\\
  P^{(n)}_r&=&\frac{(-1)^{n+1}}{2\pi i}\oint_{\Gamma_{r}}\,R_0(z)[VR_0(z)]^n\,dz, \;n\geq 1
\end{eqnarray}
whence
\begin{eqnarray}
\label{exp1}
P_r(-g)H(g)P_r(g) = \sum_{n=0}^\infty g^n  \hat {T}^{(n)}_r, \quad  \hat {T}^{(0)}_r=H_0P_r
\end{eqnarray}
where
\be
\label{coeffT}
\hat {T}^{(n)}_r = \sum_{p=0}^n (-1)^p[ P^{(p)}_rH_0 P^{(n-p)}_r +  P^{(p-1)}_rV P^{(n-p)}_r]\,,\quad n\geq 1,\quad  P^{(-1)}_r=0\,.
\ee
and \be
\label{Pn}
P^{(n)}_r=(-1)^{n+1}\sum_{{k_1+\dots +k_{n+1}=n}\atop{\,k_j \geq 0}}S^{(k_1)}_rVS^{(k_2)}_rV\dots VS^{(k_{n})}_rVS^{(k_{n+1})}_r\,.
\ee
Here
\be
\label{SSS}
S^{(0)}_r = -P_r\,;\quad S_r = -\sum_{j\neq r}P_j/(\l_j - \l_r)\,; \quad S^{(k)}_r = (S_r)^k\,, \quad \forall k=1,2,...,
 \ee
 where $P_j$ is the projection corresponding to the eigenvalue $\l_j$ of $H_0$.
 \item[(2)]
The series (\ref{exp0},\ref{exp1}) are  norm convergent  for $\ds |g|<\frac{d_r}{2\|W\|_\infty}$, where $d_r$ is the distance of $\l=\l_r$ from the rest of the spectrum of $H_0$. Hence under the present assumptions the convergence takes place a fortiori for
 \be
\label{rc}
|g|<\rho\,\qquad  \rho:=\frac{\delta}{\|W\|_\infty}. 
\ee
\item[(3)]
The projection operator $P_r(g)$ is holomorphic  for $|g|<\rho$. This entails that its dimension is constant throughout the disk. Therefore $H(g)$ admits exactly $m_r$ eigenvalues $\l_{r,s}$ (counting multiplicities) inside $\Gamma_r$ for $|g|<\rho$.  
\item[(4)] 
Hence, for $|g|<\rho$ we can write:
\be
\label{Gn}
B_r(g)=\sum_{n=0}^\infty g^n{\cal G}^{(n)}_r, \quad ({\cal G}^{(n)}_r)_{jl}:= \la\psi_{r,j}, \hat {T}^{(n)}_r\psi_{r,l}\ra, \quad j,l=1,\ldots,m_r\,.
\ee
\end{itemize}
We can now formulate the first step:
\begin{proposition}
\label{propo1} 
Let $\l_r, r=1,2,\ldots$ be an eigenvalue of $H_0$. Then    the $m_r$ eigenvalues (counting multiplicity) $\l_{r,s}$ of $H(g)$ existing  for $|g|<\rho$, and converging to $\l_r$  as $g\to 0$, are  real  for $|g|<\bar g(r)$, $ g\in\R$.
\end{proposition}
{\bf Proof}
\newline
 We drop the index $r$ because the argument is $r-$independent, i.e. we consider the expansion near the unperturbed eigenvalue $\l:=\l_r$.  Accordingly, we denote by $\psi_s:=\psi_{r,s}$ the corresponding eigenvectors.  Let us first consider the case of $\l$ even.
It is enough to prove that ${\cal G}^{n}=0$ if $n$ is odd and that ${\cal G}^{n}$ is selfadjoint (in fact, real symmetric) when $n$ is even. These assertions will be proved in Lemma \ref{zero} and  \ref{simmetria}, respectively, which in turn require   an auxiliary statement.
\begin{definition}
\label{string}
The product
\be
\label{stringa}
\Pi( k_1,\ldots,k_{n+1}):=S^{(k_1)}VS^{(k_2)}V\dots VS^{(k_{n})}VS^{(k_{n+1})}
\ee
containing precisely $n$ factors $V$ and $n+1$ factors $S^{(j)}$, $j\geq 0$,  is called {\rm string of length } $n$.
\end{definition}
Then from (\ref{coeffT},\ref{Pn}) we get:\be
\label{G1}
 ({\cal G}^{(n)})_{qs} =  (-1)^n\sum_{p=0}^n(-1)^p[({\cal G}^{(n)}_{1,p})_{qs} - ({\cal G}^{(n)}_{2,p})_{qs} ]
 \ee
 where
 \bea
  &&({\cal G}^{(n)}_{1,p})_{qs} =
 \langle \psi_q,\sum_{{k_1+\dots +k_{p+1}=p}\atop{\,k_l \geq 0}}\Pi(k_1,\ldots,k_{p+1}) H_0\sum_{{h_1+\dots +h_{n-p+1}\atop{=n-p};\,h_l \geq 0}}\Pi(h_1,\ldots,h_{n-p+1})\psi_s\ra \qquad\quad
\label{G2}\\
&& ({\cal G}^{(n)}_{2,p})_{qs} = \langle \psi_q,\sum_{{k_1+\dots +k_p}\atop{=p-1;\,k_l \geq 0}}\Pi(k_1,\ldots,k_p)V\sum_{{h_1+\dots +h_{n-p+1}}\atop{=n-p;\,h_l\geq 0}}\Pi(h_1,\ldots,h_{n-p+1})\psi_s\rangle  \qquad\quad 
\label{G3}
\eea
Now  $S^{(k)}$ is selfadjoint for all $k$, and $V=iW$ with $W(x)\in\R$.  Therefore:
\begin{eqnarray}
&&({\cal G}^{(n)}_{1,p})_{qs} =(-1)^p
 \langle \sum_{{k_1+\dots +k_{p+1}=p; \; k_l\geq 0} \atop{\,h_1+\dots +h_{n-p+1}=n-p;\; h_l\geq 0}}\Pi(k_{p+1},\ldots,k_{1}) \psi_q,H_0\Pi(h_1,\ldots,h_{n-p+1})\psi_s\ra
\nonumber\\
&& ({\cal G}^{(n)}_{2,p})_{qs} =(-1)^{p-1} \langle \sum_{{k_1+\dots +k_p=p-1;\; k_l\geq 0} \atop{\,h_1+\dots +h_{n-p+1}=n-p; \; h_l\geq 0}}\Pi(k_p,\ldots,k_1)\psi_q,
V\Pi(h_1,\ldots,h_{n-p+1})\psi_s\rangle \,.
\nonumber
\end{eqnarray}
Since $S^{(k)}\perp P$, $k\geq 1$, in both scalar products (\ref{G2})
and (\ref{G3}) all terms  with $k_1\neq 0$ or $h_{n-p+1}\neq 0$ vanish. Hence:
 \bea
 \label{G4}
 && ({\cal G}^{(n)}_{1,p})_{qs}=(-1)^{p}\langle \sum_{{k_1+\dots +k_{p}=p}\atop {{h_1+\dots +h_{n-p}=n-p}\atop{k_l\geq 0, h_l\geq 0}}}\Pi(k_p,\ldots,k_{1})V\psi_q, H_0\Pi(h_1,\ldots,h_{n-p})V\psi_s\ra \qquad\quad
 \\
 \label{G5}
 && ({\cal G}^{(n)}_{1,p})_{qs}=(-1)^{p-1}\langle \sum_{{k_1+\dots
     +k_{p-1}=p-1}\atop {{h_1+\dots +h_{n-p}=n-p}\atop{k_l\geq 0,
       h_l\geq 0}}}\Pi(k_{p-1},
\ldots,k_{1})V\psi_q, V\Pi(h_1,\ldots,h_{n-p})V\psi_s\ra \qquad\quad
 \eea 
 We now have:
\begin{lemma}
\label{zero}
Let $n$ be odd, and $0\leq p\leq n$. Then, $\forall\,k_1,\ldots,k_p\geq 0$, $\forall\,h_1,\ldots,h_{n-p}\geq 0$, $\forall\,q,s=1,\ldots,m$:
\begin{eqnarray}
\label{zero1}
\la \Pi(k_{p},\ldots,k_1)V\psi_q,H_0\Pi( h_1,\ldots,h_{n-p})V\psi_s\ra=0
\\
\label{zero2}
\la \Pi(k_{p-1},\ldots,k_1)V\psi_q,V\Pi( h_1,\ldots,h_{n-p})V\psi_s\ra=0
\end{eqnarray}
\end{lemma}
{\bf Proof}
\newline
 Let us write explicitly   (\ref{zero1},
 \ref{zero2}):
 \begin{eqnarray}
 \label{zero3}
\la S^{(k_{p})}VS^{(k_{p-1})}V\dots VS^{(k_{1})}V\psi_q,H_0S^{(h_1)}VS^{(h_2)}V\dots VS^{(h_{n-p})}V\psi_s\ra=0
\\
\label{zero4}
\la S^{(k_{p-1})}VS^{(k_{p-2})}V\dots VS^{(k_{1})}V\psi_q,VS^{(h_1)}VS^{(h_2)}V\dots VS^{(h_{n-p})}V\psi_s\ra=0
\end{eqnarray} 
Let us now further simplify the notation as follows. We set:
\be
\label{S+S-}
S_+ :=  -\sum_{j\neq r;\, \l_j {\rm even}}\frac{P_j}{\l_j - \l}\,; \quad S_- =  -\sum_{j\neq r;\, \l_j {\rm odd}}\frac{P_j}{\l_j - \l}\,.
\ee
Both series are convergent because $|(\l_j - \l)|>\delta$ and $\ds \sum_{j\neq r;}P_j$ is convergent.   Hence  $S= S_+ \oplus  S_-$  and  for $k\neq 0$ we have:
\be
\label{Powers}
S^k = S_+^k \oplus S_-^k = (-1)^k \sum_{j\neq r;\, \l_j {\rm even}}\frac{P_j}{(\l_j - \l)^k }+ (-1)^k \sum_{j\neq r;\, \l_j {\rm odd}}\frac{P_j}{(\l_j - \l)^k}\,.
\ee
Finally we set $S_+^{(0)}:=S_+^0:=-P$. Now, the multiplication by $V$ changes the parity of a function, and $\psi_j, \psi_l$ are even. This entails that in both scalar products above  $S^{(k_1)}$ can be replaced by  $S_-^{(k_1)}$,  $S^{(k_2)}$ by  $S_+^{(k_2)}$ and so on. The general rule is: $S^{(k_j)}$ can be replaced by $S_-^{k_j}$ ( by $S_+^{k_j}$) if and only if $j$ is odd ($j$ is even, respectively). 
Similarly for the $S^{(h_j)}$. Consider first the scalar product in (\ref{zero3}). According to the general rule $S_{\pm}^{(k_p)}$ coincides with $S_+^{(k_p)}$ if $p$ is even and with $S_-^{(k_p)}$ if $p$ is odd. Similarly for 
$S_{\pm}^{(h_{n-p})}$. If $n$ is odd $p$ and $n-p$ have opposite parity and since $H_0$ does not change the parity of a function the scalar product is zero. A similar argument shows that also the scalar product (\ref{zero4}) is zero if $n$ is odd. Indeed the function in the left hand side has the same parity of the number $p-1$, whereas the function of the right hand side has the same parity of $n-p+1$, and if $n$ is odd $p-1$ and $n-p+1$ have opposite parity. This proves the assertion.
\begin{lemma}
\label{pari} Let $n$ be odd. Then ${\cal G}^{(n)}=0$.
\end{lemma}
{\bf Proof}
\newline
It is an immediate consequence of Lemma \ref{zero} on account of (\ref{G1},\ref{G4},\ref{G5}). 
\begin{lemma}
\label{simmetria}
 Let $n$ be even. Then
$({\cal G}^{(n)})_{qs} = \overline{({\cal G}^{(n)})_{qs}}$ for all $q,s = 1,\dots,m$. 
\end{lemma}
{\bf Proof}
\newline
Once more by (\ref{G1},\ref{G4},\ref{G5})  we can write  for all $n$ (replacing of course $V$ by $iW$ in the definition (\ref{stringa}), and denoting $\Pi^\prime$ the resulting string)
\begin{eqnarray}
\label{coeffG3}
 &&({\cal G}^{(n)})_{qs} =
 \nonumber\\
 &&(i)^n\sum_{p=0}^n
[\sum_{{k_1+\dots +k_p=p; k_j\geq 0} \atop{\,h_1+\dots +h_{n-p}=n-p;\; h_j\geq 0}}(-1)^p\langle \Pi^\prime(k_1,\dots ,k_p)W\psi_q,
H_0\Pi^\prime(h_1,\ldots,h_{n-p})W\psi_s\rangle 
\nonumber\\
&& -\sum_{{k_1+\dots +k_{p-1}=p-1; k_j\geq 0} \atop{h_1+\dots +h_{n-p}=n-p; \; h_j\geq 0}}(-1)^{p-1}\langle \Pi^\prime(k_1,\dots ,k_{p-1})W\psi_q,
W\Pi^\prime(h_1,\ldots,h_{n-p})W\psi_s\rangle ]=
\nonumber\\
&&(i)^n\sum_{p=0}^n
[\sum_{{k_1+\dots +k_{n-p}=n-p;} \atop{h_1+\dots +h_p=p; h_j\geq 0, k_j\geq 0}}(-1)^{n-p}\langle H_0\Pi^\prime(k_1,\ldots,k_{n-p})W\psi_q,
\Pi^\prime(h_1,\ldots,h_p)W\psi_s\rangle 
\nonumber\\
&&  -\sum_{{k_1+\dots +k_{n-p}=n-p; k_j\geq 0} \atop{\,h_1+\dots +h_{p-1}=p-1; 
 h_j\geq 0}}(-1)^{n-p+1}\langle W\Pi^\prime(k_1,\ldots,k_{n-p})W\psi_q,
\Pi^\prime(h_1),\ldots,h_{p-1})W\psi_s\rangle ] 
\nonumber\\
&&= \overline{({\cal G}^{(n)})_{qs}}\,.
\end{eqnarray}
To obtain the second equality in (\ref{coeffG3}) we have used the selfadjointness of $H_0$ and $W$ and we have renamed the indices, exchanging $p$ and $n-p$ in the first scalar product, and $p-1$ and $n-p$ in the second scalar product. Finally, to obtain the last equality in (\ref{coeffG3}) notice that $(-1)^p = (-1)^{n-p}$ since $n$ is even.
\par\noindent
{\bf Remarks} 
\begin{enumerate}
\item
It is worth noticing that if the $\psi_s , s=1,\dots,m$, are chosen to be real valued then $({\cal G}^{(n)})_{qs}\in\R, \forall j,l$, because $W$ is also real valued and the operators $S^{(k)}$ map real valued functions into real valued functions.
\item
The argument yielding the real nature of the perturbation expansion is independent of its convergence, namely it holds for all odd potentials $V$ for which the perturbation expansion exists to all orders. In particular, it holds when $V$ is any odd polynomial, i.e. for any odd anharmonic oscillators in any dimension $d$.
\end{enumerate}
We now proceed to prove that the eigenvalues $\ell_{r,s}(g)$ are real $\forall\,g\in\R$, $|g|<\rho$. 
\begin{proposition}
\label{anglobale}
The eigenvalues $\l_{r,s}$, $r=1,2,\ldots$, $s=1,\ldots,m_r$   are holomorphic for $|g|<\rho$ and real for $g\in\R$,  $|g|<\rho$. 
\end{proposition}
{\bf Proof}
\newline
The vectors  $U_r(g)P_r\psi_{r,k}=U_r(g)\psi_{r,k}: k=1,\ldots,m_r$ represent a basis of ${\cal M}_r(g)$ for all $|g|<\rho$ (\cite{Ka}, \S II.4.2).  Here the similarity operator $U_r(g)P_r$ is recursively defined in the following way:
\be
\label{U(g)}
U_r(g)P_r=P_r+\sum_{k=1}^\infty U^{(k)}_rg^k, \quad kU^{(k)}_r=kP^{(k)}_r+(k-1)P^{(k-1)}_rU^{(1)}_r +\ldots P^{(1)}_rU^{(k-1)}_r
\ee
We denote $\chi_{r,s}(g): s=1,\ldots,m_r$ the orthonormal basis in ${\cal M}_r(g)$ obtained from  $U_r(g)\psi_{r,s}$, $s=1,\ldots,m_r$ through the Gram-Schmidt orthogonalization procedure. Then the eigenvalues  $\l_{r,s}(g), s=1,...,m_r$, are the eigenvalues of the $m\times m$ matrix $X_{r}(g)$ given by:
\begin{eqnarray}
\nonumber
(X_{r}(g))_{hk}&:=&\la\chi_{r,h}(g), H(g)P(g)\chi_{r,k}(g)\ra
\\
&=&\la\chi_{r,h}(g), H(g)\chi_{r,k}(g)\ra,\quad h,k=1,\ldots,m_r\,.
\end{eqnarray}
because $P(g)\chi_{r,h}(g)=\chi_{r,h}$, $h=1,\ldots,m_r$.  For $|g|<\bar g(r)$ the orthonormal  vectors $\chi_{r,h}(g): h=1,\ldots,m_r$ are linear combinations of   the orthonormal vectors $\phi_{r,h}(g): h=1,\ldots,m_r$ defined above.  Since   $X_{r}(g)$ and $T_{r}(g)$ represent the same operator on two different orthonormal basis, if either one is self-adjoint the second must enjoy the same property. Hence  the matrix $(X_{r}(g))_{hk}$ is self-adjoint, $|g|<\bar g(r)$, $g\in\R$. Expand now $(X_{r}(g))_{hk}$ in power series:
$$
(X_{r}(g))_{hk}=\sum_{m=0}^\infty(\theta_{r,m})_{hk}g^m
$$
The series converges for $|g|<\rho$. It follows  indeed by the standard Gram-Schmidt procedure (we omit the details) that it can  be written as the quotient of two functions of $g$ involving only linear combinations of scalar products of the operators $P_r(g)$ on vectors independent of $g$; the denominator never vanishes for $|g|<\rho$ by construction, on account of the linear independence of the vectors $U_r(g)\psi_{r,s}$, $s=1,\ldots,m_r$ when $|g|<\rho$.  Now it  necessarily follows from the self-adjointness  of $(X_{r}(g))_{hk}$, valid for $|g|<\bar g(r)$  that $(\theta_{r,m})_{hk}=\overline{(\theta_{r,m})}_{kh}$, $m=0,1,\ldots$. 
 Hence the matrix  $X_{r}(g)$ is self-adjoint for $|g|<\rho$, $g\in\R$, and thus the eigenvalues $\l_{r,s}$ are real in the same domain. This proves the assertion.
\vskip 0.3cm\noindent
{\bf Proof of Theorem \ref{thm2}}
\newline
We have seen that  the RSPE associated with the $\l_r$-group of eigenvalues $\l_{r,s}(g), s=1,\dots, m_r$, of $H(g)$ which converge to $\l_r$ as $g\to 0$, have radius of convergence no smaller than $\rho$. Hence, $\forall g\in \R$ such that $|g|<\rho$, $H(g)$ admits a sequence of real eigenvalues $\l_{r,s}(g), s=1,\dots, m_r, r\in \N$.  We want to prove that for $|g|<\rho, g\in\R$, $H(g)$ has no other eigenvalues.  Thus all its eigenvalues are real. To this end, for any $r\in\N$  let ${\cal Q}_r$ denote the square centered at $\l_r$ with side $2\delta$. 
 Then if $g\in\R$, $|g|<\rho$, and $\l(g)$ is an eigenvalue of $H(g)$:
$$
\l(g)\in\bigcup_{r\in\N}\,{\cal Q}_r .
$$
In fact, for any $\ds z\notin \cup_{r\in\N}\,{\cal Q}_r$ we have
\be
\label{Stima1}
\|gVR_0(z)\|\leq |g|\|W\|_\infty \|R_0(z)\|<\rho \|W\|_\infty[{\rm dist}(z,\sigma(H_0))]^{-1}\leq \frac{\rho \|W\|_\infty}{\delta}= 1
\ee
where $R_0(z):= (H_0 - z)^{-1}$. Thus, $z\in\rho (H(g))$ and 
$$
R(g,z):= (H(g) - z)^{-1} = R_0(z)[1 + gVR_0(z)]^{-1}\,.
$$
Now let $g_0\in\R$ be fixed with $|g|<\rho $. Without loss of generality we assume that $g_0>0$.  Let $\l(g_0)$ be a given eigenvalue of $H(g_0)$. Then  $\l(g_0)$ must be contained in the interior (and not on the boundary) of ${\cal Q}_{n_0}$ for some $n_0\in\N$. Moreover if $m_0$ is the multiplicity of $\l(g_0)$, for $g$ close to $g_0$ there are $m_0$ eigenvalues $\l^{(\alpha)}(g), \alpha=1.\dots ,m_0$, of $H(g)$ which converge to $\l(g_0)$ as $g\to g_0$ and each function $\l^{(\alpha)}(g)$ represents a branch of one or several holomorphic functions which have at most algebraic singularities at $g=g_0$ (see [Kato, Thm. VII.1.8]). Let us now follow one of such branches $\l^{(\alpha)}(g)$ for $0<g<g_0$, suppressing the index $\alpha$ from now on. First of all we notice that, by continuity, $\l(g)$ cannot go out of ${\cal Q}_{n_0}$ for $g$ close to $g_0$. Moreover, if we denote $\Gamma_{2t}$ the boundary of the square centered at $\l_{n_0}$ with side $2t$, for $0<t\leq 1$, we have, for $z\in\Gamma_{2t}$ and $0<g\leq g_0$,
\be
\label{Stima2}
\|gVR_0(z)\|\leq g[{\rm dist}(z,\sigma(H_0))]^{-1}\leq g/t \,.
\ee
Then $t>g$ implies $z\notin \sigma(H(g))$, i.e. if $z\in \sigma(H(g))\cap\Gamma_{2t}$ then $t\leq g<g_0<1$. Hence we observe that as $g\to g_0^-$, $\l(g)$ is contained in the square centered at $\l_{n_0}$ and side $2g$. Suppose that the holomorphic function $\l(g)$ is defined on the interval $]g_1, g_0]$ with $g_1>0$. We will show that it can be continued up to $g=0$, and in fact up to $g=-1$. From what has been established so far the function $\l(g)$ is bounded as $g\to g_1^+$. Thus, by the well known properties on the stability of the eigenvalues of the analytic families of operators, $\l(g)$ must converge to an eigenvalue $\l(g_1)$ of $H(g_1)$ as $g\to g_1^+$ and $\l(g_1)$ is contained in the square centered at $\l_{n_0}$ and side $2g_1$. Repeating the argument starting now from $\l(g_1)$, we can continue $\l(g)$ to a holomorphic function on an interval $]g_2, g_1]$, which has at most an algebraic singularity at $g=g_2$. We build in this way a sequence $g_1>g_2>\dots >g_n>\dots $ which can accumulate only at $g=-1$. In particular the function $\l(g)$ is piecewise holomorphic on  $]-1, 1]$. But while passing through $g=0$, $\l(g)$ coincides with one of the eigenvalues $\l_{r,s}(g), s=1,\dots, m_r$, generated by an unperturbed eigenvalue $\l_r$ of $H_0$ (namely $\l_{n_0}$), which represent $m_r$ real analytic functions defined for $g\in ]-1,1]$. Thus, $\l(g_0)$ arises from one of these functions and is therefore real. This concludes the proof of the Theorem. 
\vskip 3pt\noindent
\section{Perturbation of resonant harmonic oscillators}
\setcounter{equation}{0}%
\setcounter{theorem}{0}%
Consider  again the $d$-dimensional harmonic oscillator  
\be
\label{H_00}
H_0=\frac12\sum_{k=1}^d\left[-\frac{d^2}{dx_k^2}+\om_k^2x^2_k\right]
\ee
where now the frequencies $\om_k>0: k=1,\ldots,d$ may be different.  Theorem \ref{thm3} will be a consequence of the following
\begin{proposition}
\label{prop2}
The operator (\ref{H_00}) fulfills Assumption (2) of Theorem \ref{thm2} if and only if the following condition on the frequencies holds:
\begin{itemize}
\item[(A)]
$\forall\,k\in\Z^d\setminus\{0\}$ such that the components $k_i: i=1,\ldots,d$ have no common divisor, and  $\om_1k_1+\ldots+\om_dk_d=0$, the number $O(k)$ of $k_i$ odd is even.
\end{itemize}
\end{proposition}
{\bf Proof}
\newline
We first prove the sufficiency part. Let therefore (A) be fulfilled.  First recall the obvious fact that the rational dependence of the frequency entails the degeneracy of any eigenvalue of $(\ref{H_00})$.  In order to show  that   each eigenvalue
 $$
\l_{n_1,\ldots,n_d}=\om_1n_1+\ldots\om_dn_d+\frac12(\om_1+\ldots\om_d)
$$
of $H_0$ has a definite parity,   consider a corresponding  eigenfunction
$$
\Psi_{n_1,\ldots,n_d}(x_1,\ldots,x_d)=\prod_{s=1}^d\,\psi_{n_s}(x_s)
$$
Now $\psi_{n_s}(x)$ is even or odd according to the parity of $n_s$,
and therefore $\Psi$ will be even if and only if  the number of odd
$n_s$ is even.    Since $\l$ is degenerate, there  exist
$(l_1,\ldots,l_d)\neq (n_1,\ldots,n_d)$ such
that
$$
\om_1n_1+\ldots +\om_dn_d=\om_1l_1+\ldots+\om_dl_d \Longrightarrow \la\om,k\ra=0, \;k:=(n_1-l_1,\ldots,n_d-l_d)
$$
and hence the eigenfunction 
$$
\Psi_{l_1,\ldots,l_d}(x_1,\ldots,x_d)=\prod_{s=1}^d\,\psi_{l_s}(x_s)
$$
corresponds to the same eigenvalue.  The eigenfunctions $\Psi_{n_1,\ldots,n_d}$ and $\Psi_{l_1,\ldots,l_d}$ have one and the same parity if and only if the number of the odd differences $k_i$ is even: in fact, an even difference $k_i=n_i-l_i$ does not change the relative parity, while an odd difference does.  Let us show that  if Assumption (A) holds  the number of odd differences is even. The case in which  $k_i: i=1,\ldots,d$ have no common divisor is the Assumption itself.   Let therefore $k_i: i=1,\ldots,d$ have a common divisor. If a common divisor is $2$,  $k_i$ is even for any $i$. Hence there are no odd differences. If $2$ is not a common divisor, there will be an odd common divisor, denoted $b$, such that $k_i=bk^\prime_i$, where the numbers $k^\prime_i$ have no common divisor. Now $\la k^\prime,\om\ra=\la k,\om\ra/b=0$. Hence by the assumptions  $O(k^\prime)$ is  even. Since the multiplication by the odd number $b$ does not change the parity of the $k^\prime_i$, the same conclusion applies also to the numbers $k_i$. Thus the total number of odd differences does not change after multlplication by $b$: $O(k)=O(k^\prime)$ is even.
\par\noindent
Conversely, let us assume that Assumption (A) is violated. Therefore there exists $ k\in 
\Z^d\setminus\{0\}$ 
such that the numbers $k_i$ have no common divisor, $\la k,\om\ra=0$ and  $O(k)$ is odd.  Consider again the eigenfunctions $\Psi_{n_1,\ldots,n_d}(x_1,\ldots,x_d)$ and $\Psi_{l_1,\ldots,l_d}(x_1,\ldots,x_d)$ corresponding to the same eigenvalue $\l$, with $k_i=n_i-l_i$ as above.  By construction, the two eigenfunctions have opposite parity, and this concludes the proof of the Proposition.
\par\noindent
{\bf Proof of Theorem \ref{thm3}}
\newline
Let us first prove that Assumption (1) of Theorem \ref{thm2}  is fulfilled.  Let $\l_l=\l_{l_1,\ldots,l_d}$ and $\l_n=\l_{n_1,\ldots,n_d}$ denote different eigenvalues. Then, by assumption:
\begin{eqnarray*}
|\l_n-\l_l|=\om |(n_1-l_1)\frac{p_1}{q_1}+\ldots+(n_d-l_d)\frac{p_d}{q_d}|=\qquad
\\
\frac{\om}{q_1\cdots q_d}|(n_1-l_1)p_1q_2\cdots q_d+\ldots+(n_d-l_d)p_dq_1\cdots q_{d-1}|
\\
\geq  \frac{\om}{q_1\cdots q_d}:=\delta>0\qquad\qquad\qquad\qquad\qquad
\end{eqnarray*}
Since this lower bound does not depend on the multi-indices $(n,l)$ the assertion is proved. 
\newline
Let us now check Assertion  (i), namely that if the frequencies have the form 
$\om_k=\om p_k/q_k$ with $p_k$ and $p_k$ odd then Assertion  (2) of Theorem (\ref{thm2}) holds;
 namely, all eigenvalues of (\ref{H_00}) have a definite parity.  By Proposition 4.2, it is enough to prove that  Assumption (A) is satisfied. Let indeed 
 $(k_1,\ldots,k_d)\in\Z^d\setminus\{0\}$ be without common divisor and such that $\la \om,k\ra=0$. Then: 
\begin{eqnarray*}
\frac{p_1}{q_1}k_1+\ldots+\frac{p_d}{q_d}k_d&=
&\frac{1}{q_1\cdots q_d}(p_1q_2\cdots q_dk_1+\ldots+p_dq_1\cdots q_{d-1}k_d)=
\\
&:=&\frac{1}{q_1\cdots q_d}(D_1k_1+\ldots+D_dk_d)=0
\end{eqnarray*}
Now the integers $D_k: k=1,\ldots,d$ are odd; hence the above sum must have an even number of terms. 
The odd terms are those, and only those, containing an odd $k_i$; 
therefore the number of odd $k_i$ must be even. Then the result follows by the above Proposition. 
\newline
Consider now Assertion (ii) of Theorem \ref{thm3}.  
The only thing left to prove is that the validity of Assumption (A) entails that 
$\ds \frac{\om_1}{\om_2}=\frac{d_1}{d_2}$ where $d_1$ and $d_2$ are odd.  
Suppose indeed $\ds \frac{\om_1}{\om_2}=\frac{k_2}{k_1}$ where $k_1$ is odd and $k_2$ even, or viceversa. 
Then $\om_1k_1-\om_2k_2=0$- However this contradicts 
Assumption (A) which states that the number $O(k)$ of odd $k_i$ must be even. 
This concludes the proof of Theorem \ref{thm3}.
\begin{corollary}
\label{cor1} Under the conditions of Theorem  \ref{thm3} on $H_0$, assume furthermore that the matrix $\la\psi_r,W\psi_s\ra: r,s=1,\ldots,m_0 $ is not identically zero for at least one eigenvalue $\l_0$ of $H_0$ of  multiplicity  $m_0>1$. Then  for  $\ds |g|<\frac{\delta}{\|W\|_\infty}$, 
 $H(g)$ has  real eigenvalues if and only if $p$ and $q$ are both odd.
\end{corollary}
{\bf Proof}
\newline
The sufficiency part is a particular case of Theorem \ref{thm3}.  As for the necessity, under the present conditions the eigenfunctions have opposite parity. Thefore we can directly apply the argument of \cite{CGS} and conclude that if $p$ is even and $q$ odd or  viceversa $H(g)$ has a pair of complex conjugate eigenvalues near $\l_0$ for $g\in\R$ suitably small.

\end{document}